\documentstyle[prl,aps]{revtex}

\begin{document}
\twocolumn[{
\draft
\author{B.A.Muzykantskii and D.E.Khmelnitskii}
\address{Cavendish Laboratory, University of Cambridge,
Madingley Road, Cambridge CB3 0HE, UK}
\address{and L.D.Landau Institute for Theoretical Physics, Moscow,
Russia}

\title{Nearly Localized States in Weakly Disordered Conductors}
\date{\today}
\maketitle
\widetext
\begin{abstract}
\leftskip 54.8pt
\rightskip 54.8pt
The time dispersion of the averaged conductance $G(t)$ of a mesoscopic
sample is calculated in the long time limit when  $t$  is much larger
than the diffusion travelling time $ t_D$. In this case the functional
integral in the effective supersymmetric field theory is determined by
the saddle point contribution.  If  $t$ is shorter than the inverse
level spacing $\Delta$ ($\Delta t / \hbar \ll 1$), then $G(t)$ decays
as $\exp[-t/t_D]$.  In the ultra-long time limit ($\Delta t / \hbar \gg
1$) the  conductance $G(t)$ is determined by the electron states that  are
poorly connected with the outside leads. The probability to find such a
state decreases more slowly  than any exponential funcion as $t$ tends to
infinity.
It is worth mentioning, that the saddle point equation looks very
similar to the well known Eilenberger equation in the theory of dirty
superconductors.
\end{abstract}
\leftskip 54.8pt
\pacs{PACS numbers: 73.40.Gk, 05.45.+b,  73.20.Dx, 72.20.My}
}]

\narrowtext

We consider long time relaxation phenomena in a disordered conductor
that is attached to ideal leads.
For simplicity we assume that the electrons in this conductor
do not interact with each other, the temperature is zero ($T=0$)
and there are no inelastic processes.
The total current $I(t)$ at time $t$ depends
upon the voltage according to the Ohm law:
$$
 I(t) = \int_{-\infty}^{t}dt' G(t-t') V(t').
$$
We are interested in the asymptotic form  of the conductance $G(t)$
as $ t \to \infty$.

The same problem has been   considered  earlier by Altshuler, Kravtsov and
Lerner (AKL)  \cite{Alt-Kra-Ler}.
Our initial goal was to obtain their results  by means of a
more direct calculation. At present, we can neither confirm,
nor disprove the AKL results. We have found an
intermediate range of times, where the conductance $G(t)$ decays more slowly
than it was predicted in Ref. \cite{Alt-Kra-Ler}. The AKL asymptote
could be valid at longer times (see discussion below).

There are three time scales in the problem:
\begin{enumerate}
\item  The mean free time $\tau =  l / v_F$, where $v_F$ is the  Fermi
       velocity  and $l$ is the mean free path. This time scale determines
       the  dispersion of the Drude conductivity
         $\sigma_0  \sim e^{-t/\tau} $
\item  The time of diffusion through the sample
       $t_D= L^2/ D$, where $D=  l^2/ 3 \tau$ is the diffusion coefficient,
        and $L$ is the sample size.
\item  The inverse mean level spacing $\hbar / \Delta = \hbar \nu V $,
       where $\nu$ is the density of states and V is the volume of
       the sample.
\end{enumerate}
In a macroscopic sample the inequality $ \tau \ll t_D \ll \hbar
/\Delta$  is valid, provided that $ L \gg l$ and the disorder is weak.
Indeed, the product $\Delta t_D $ is connected with the dimensionless
conductance of the sample $ g = 2 \pi \hbar/( t_D \Delta)$, which is
large for a weak disorder. The times $t_D$ and $\hbar/ \Delta $  enter
into time dispersion only due to quantum corrections to conductivity.

At times $ t \ll \hbar/ \Delta$ an electron can be considered  a
wave packet of many superimposed states
propagating semi-classically.
Therefore, it is natural to assume that the conductance
$G(t)$ is proportional to the probability of finding a
Brownian trajectory that remains in the sample for the time $t$.
For $ t \gg t_D$ such a probability decays as
 $ \exp \left[-t/t_D \right]$.
Our calculations confirm  this result.

In the opposite limit, for $t \gg \hbar/\Delta$, the conductance $G(t)$ is
proportional to the probability of finding an  electron state with the
life time $ t$.
In order to trap an electron for a long time the state must be poorly
connected with the leads (nearly
localized).  We show that the probability of finding such a state decays
non-exponentially with time.  Namely,
$G(t) \sim \exp \left[-g \log^2 ( t
\Delta ) \right] $ for $d=1$ and $ G(t) \sim ( t \Delta
)^{-g} $ for $d=2$.
These results are not valid in the very long time limit.
We discuss this later together with the question of dimensional
crossover.

Instead of calculating the  conductance as a function of time, we could have
worked in the frequency representation. In that way we would have found
a singularity in $ G (\omega) $ as $ \omega \to 0$.  This singularity,
however, does not affect the value of the d.c.  conductance and therefore has
an obscure physical meaning, while the time domain results
have the direct interpretation.

Since the long time asymptote corresponds to the rare events
when the electron is nearly trapped in the  sample, it is natural to use
the saddle-point approximation.  We carry out the following
program:
\begin{enumerate}
\item Express the averaged conductance
\cite{conductance fluctuations}
as a functional integral over
  supermatrices
       \cite{prefactor}
    (see \cite{Efetov-review,Weidenmuller} for review):
    \begin{eqnarray}
&& G(t) = G_0 e^{-t/\tau} + \nonumber \\
&& \int \frac{d \omega}{2 \pi} e^{-i \omega t}
 \int {\cal D} Q(r)  P\{Q\} \exp \left[-A \right],
\label{G(t)}  \\
&& A = \frac{\pi \nu}{8} \int dr  \;
         \mbox{Str} \{ D (\nabla Q)^2  + 2 i \omega \Lambda Q \},
   \nonumber
    \end{eqnarray}
\item
      Vary the action $A$ with respect to $Q$, taking into account
      the constraint $Q^2=1$,  and obtain the saddle-point condition which
recalls the diffusion limit of the Eilenberger equation
\cite{Eilenberger}:
     \begin{equation}
      2 D \nabla ( Q \nabla Q) + i \omega \left[ \Lambda, Q \right] =0
     \label{saddle-point}
     \end{equation}
\item
      Derive the condition at the boundary  with the lead
      \begin{equation}
       \left. Q \right|_{\mbox{\small lead}} =  \Lambda .
        \label{boundary-cond}
      \end{equation}
\item
      Perform the integration over $ \omega$ in Eq. (\ref{G(t)}) and
      obtain the self consistency condition
       \begin{equation}
      \int \frac{dr}{V} \; \mbox{Str} ( \Lambda Q )
       = - \frac {4 t \Delta}{ \pi \hbar}
      \label{self-cons}
      \end{equation}
      which allows us to exclude  $\omega$ from Eq.~(\ref{saddle-point}).
\item
 	Substitute the solution of Eq.~(\ref{saddle-point}) with boundary
	conditions (\ref{boundary-cond}) in Eq.~(\ref{G(t)}) and obtain the
	results with exponential accuracy.
\end{enumerate}

The  $ 8 \times 8 $ supermatrix $Q$ has commutative
and anticommutative matrix elements. Since  $Q^2=1$  it
can be chosen in the form \cite{orthogonal}:
\begin{eqnarray}
&& Q = V^{-1} H V, \quad V=   \left(\begin{array}{cc} u & 0 \\ 0 & v
                                     \end{array}\right),
\nonumber \\
&& \Lambda = \left(\begin{array}{cc} 1 & 0 \\ 0 & -1
                                     \end{array}\right), \quad
H = \left(\begin{array}{cc} \cos \hat\theta&i \sin \hat\theta \\
                              -i \sin \hat\theta & - \cos \hat\theta
                                     \end{array}\right),
\label{Q,theta}
\\
&& \hat\theta = \left(\begin{array}{cccc} i \theta & 0  & 0 & 0\\
                                       0 & i \theta & 0 & 0 \\
                                       0 & 0 & - \theta_1 & - \theta_2 \\
                                       0 & 0 & -\theta_2 & - \theta_1
                                     \end{array}\right)
\nonumber
\end{eqnarray}

This decomposition allows us to present the action $A$ in the form
\begin{equation}
A = \frac{\pi \nu}{8} \int dr  \;
              \mbox{Str} \{ D (\nabla H)^2  + D M^2 +  2 i \omega \Lambda H \},
\label{A(H)}
\end{equation}
where $ M = \left[ V^{-1} \nabla V , H \right] $. The
minimum action  is reached for $V=\mbox{const}$, and
Eq.~(\ref{A(H)}) may be expressed in terms of
$\theta$-variables only:
\begin{eqnarray}
A&=&\frac{\pi \nu }{4} \int dr \{ 2 \,
    [D(\nabla \theta)^2 -2 i\omega \cosh \theta]  \\
&& + [D(\nabla \theta_+)^2 +2 i\omega \cos \theta_+] \nonumber \\
&& + [D(\nabla \theta_-)^2 +2i\omega \cos \theta_-] \}  \nonumber
\end{eqnarray}
where $\theta_{\pm} = \theta_1 \pm  \theta_2$. Consequently,
Eq.~(2) has the form:
\begin{mathletters}
\label{saddle-point-theta}
\begin{eqnarray}
&&  D \nabla^2  \theta + i \omega \sinh \theta =0
  \label{saddle-point-theta-a}, \\
&&  D \nabla^2  \theta_{\pm} + i \omega \sin \theta_{\pm} =0
 \label{saddle-point-theta-b}
\end{eqnarray}
\end{mathletters}
The boundary condition (\ref{boundary-cond})
follows from  the fact
that $Q$ does not fluctuate in  the bulk electrodes, $Q= \Lambda$. Hence,
at the boundary with the ideal lead $\theta=\theta_{\pm}=0$
\cite{Kupriyanov}.
The time decay of the conductance $ G(t) \sim \exp( -i \omega t) $
corresponds to real and positive values of $i \omega$.
The permitted values of frequency $\omega$ in Eq.~(\ref{saddle-point-theta-b})
are bounded from below by the value $\omega_1 \sim 1/t_D$,  which corresponds
to the linearized form of Eq.~(\ref{saddle-point-theta-b}). For smaller
frequencies
$\omega < \omega_1$, which will turn out to be the only  relevant ones,
Eq.~(\ref{saddle-point-theta-b}) has only trivial solutions $\theta _\pm =0$.
Thus, the self-consistency equation (\ref{self-cons})
has the form:
\begin{equation}
\int  \frac{d r}{V}
 \{\cosh \theta \ -1 \}  =
  \frac{t \Delta}{\pi \hbar}.
\label{self-cons-theta}
\end{equation}
The solutions of Eq.~(\ref{saddle-point-theta-a}) depend on the sample
geometry.  We start by  considering a  one dimensional wire of
length $L$,  attached to  ideal leads at $ x = \pm L/2$.
If $ t \Delta \ll 1$, then, to satisfy the self-consistency
condition (\ref{self-cons-theta}) we must choose $ \theta \ll 1$.
Therefore, Eq. (\ref{saddle-point-theta-a}) can  be  linearized. The solutions
that satisfy the boundary conditions is
\begin{equation}
 \theta = C \cos (\pi n   x /L ), \quad \omega_n =
 \frac{-i \pi^{2} n^{2}}{t_D},
 \label{cos}
\end{equation}
where $n$  is an arbitrary integer.
 The above formula for the frequency implies  that in the discussed
 regime\begin{equation}
G(t) \sim e^{-i\omega_{1} t} = \exp \left( - \frac{ \pi^{2} t }{t_D} \right).
\label{exp-decay-1d}
\end{equation}
To obtain  this result  we determine  the amplitude $C$  from
the linearized self-consistency equation, and then substitute (\ref{cos}) into
the action $A$.

For arbitrary times Eqs.~(\ref{saddle-point-theta-a}) and
(\ref{self-cons-theta}) in dimensionless coordinates have the form:
\begin{eqnarray}
&&  \frac{d^2 \theta}{d z^2} + \frac{\gamma^2}{2} \sinh \theta =0,
\quad z= \frac{x}{L},
\label{saddle-point-1d}\\
&& \int_{-1/2}^{1/2} dz [ \cosh \theta -1 ] = \frac{\Delta t}{ \pi \hbar},
 \quad \gamma^2 = 2 i \omega t_D.
\label{self-cons-1d}
\end{eqnarray}
The solution of (\ref{saddle-point-1d}) is symmetric $ \theta(z) =
\theta(-z) $, and in the region $z>0$  is given by the quadrature:
\begin{eqnarray}
&& z= \frac{1}{\gamma} \int_{\theta(z)}^{\theta_0}
 \frac{d \theta' }{ \sqrt{ \cosh \theta_0  - \cosh \theta'  } }
\label{vartheta}
\\
&& \theta_0 = \theta(0) = 2\log \frac{1}{\gamma} + 2 \log \log
\frac{1}{\gamma},
\hbox{ for } \gamma \ll 1
\label{vartheta0}
\end{eqnarray}

The function $\theta(z)$ is almost linear
$ \theta = \theta_0 (1 - 2 |z|) $ everywhere except in the region
$ |z| < 1/\log (1/\gamma) \ll 1  $.
Substituting Eq.~(\ref{vartheta}) into Eqs.~(\ref{self-cons-theta}) and
(\ref{G(t)})  we get
\begin{equation}
 i  \omega  = \frac{2 g}{t}  \log  \frac{t \Delta}{\hbar}, \quad
G(t) \sim \exp \left[ - g \log^2 \frac{t \Delta}{\hbar} \right]
\label{answer-1d}
\end{equation}

As  mentioned earlier, the contribution from the individual nearly localized
states dominates in $G(t)$ whenever $ t \Delta /\hbar \gg 1 $.
The square modulus of the wave function for such a state $ |\Psi|^2 $
equals  $\cosh \theta$.
 As we can see, this value decays exponentially
towards the leads, where $ |\Psi(x=\pm L/2)|^2 = \cosh \theta(\pm 1/2) =1$.
Because of the latter condition, the current through the wire is equal
to unity. Therefore, the escape time $t$ is proportional to the normalization
integral. This is exactly what is stated in the self-consistency condition
(\ref{self-cons-1d}) for $ \theta \gg 1$. To summarize, the wave function is
localized in
the region $ |x| \ll \xi \ll L $ with the localization length
$\xi = L/\log(t \Delta /\hbar)$ and the probability to find such a  state
is given by Eq.(\ref{answer-1d}).
For very long times, when  $\xi$ becomes less than the
transverse size of the  sample,
the one-dimensional regime crosses over
to a two-  or three-dimensional one.

In the two-dimensional case we consider a mesoscopic disk of radius $R$
surrounded by a well conducting electrode. The Laplacian operator in
Eq.~(\ref{saddle-point-theta}) is now two-dimensional and the boundary
condition is $\theta(R)=0$ at the circumference of the
disk. It is natural to assume that the minimal action corresponds to
$\theta$ that depends on the radius only and, therefore, obeys  the equation:
\begin{equation}
\theta '' + \theta' /z + i \omega t_D \sinh  \theta = 0 , \quad \theta(1) =0
\label{saddle-point-2d}
\end{equation}
where $z= r/R$ and $t_D = R^2/D$.

For $  t \ll \hbar/\Delta$ ,
Eq. ~(\ref{saddle-point-2d}) can be linearized.
Its solution  is the Bessel function
$$
 \theta = C J_0 (\gamma z),\;
  \gamma =  \sqrt{i \omega t_D} = \mu_n,
$$
where $\mu_n$ denotes the $n$-th zero of the Bessel function. The conductance
is
\begin{equation}
G(t) \sim \exp \left( - \frac{ \mu_1^2 t }{t_D} \right),
\quad  t_D \ll t \ll \hbar/\Delta.
\label{exp-decay-2d}
\end{equation}

For a long time tail $ t \gg \hbar/\Delta$, the non-linear term
in Eq.~(\ref{saddle-point-2d}) is large
near the origin and can be neglected elsewhere.  As a result,
\begin{equation}
\theta(z) = C \log \frac{1}{z}
\label{log}
\end{equation}
for all but very small $z$. On the other hand, for  $z \ll 1$,  the parameter
$\theta$ is large and $ \sinh \theta = e^\theta /2 $.
With this approximation the solution of
Eq.~(\ref{saddle-point-2d}) can be found \cite{solution-2d-small}
having the asymptote
\begin{equation}
\theta(z) = -\theta(0) + 6\log 2 + \log \frac{4}{\gamma^2} +
   4 \log \frac{1}{z},
\label{asymptote}
\end{equation}
for $ \gamma \ll z \leq 1$. Comparing with  Eq.~(\ref{log}), we have
$ \theta(0) = 6 \log 2 + \log (4/\gamma^2) $ and  $ C=4$.
To calculate the
integral in the self-consistency equation
\begin{equation}
\frac{\Delta t}{2 \pi \hbar} = \int_0^1  \{ \cosh \theta(z) -1 \} z dz
\label{self-cons-2d}
\end{equation}
we multiply Eq.~(\ref{saddle-point-2d}) by $z$, integrate in the
limits $0$ and $1$, and obtain
\begin{equation}
\left. z \frac{d \theta}{d z} \right|_0^1 +
i \omega t_D \int_0^1 \sinh \theta(z) z dz =0.
\label{trick}
\end{equation}
Since $ \theta(0) \gg 1$, we neglect the difference between the integrals in
Eqs.~(\ref{self-cons-2d}) and (\ref{trick}),
and with asymptote (\ref{log})
finally get $ i \omega = 4 g /t $.
The action $A$ is dominated by the contribution of the tail (\ref{log}):
\begin{equation}
A= 4 g \log \frac{t \Delta}{2 \pi \hbar}, \qquad
G \sim \left( \frac{\hbar}{\Delta t} \right) ^{4 g}.
\label{answer-2d}
\end{equation}
The characteristic size of the averaged 2D wave function is $\xi = \gamma R=
R (\hbar /t\Delta)^{1/2} $. The crossover to a 3D case occurs when $\xi$
becomes comparable with the film thickness.

The consideration of the 3D case makes relevant the question
of the validity of the diffusion approximation. As before, we
consider a disordered drop of radius $R$ surrounded
by a well conducting lead. Analogously to what has been
done in the 2D case, the function  $\theta$ depends
on the radius $r$ only and obeys Eq.~(\ref{saddle-point-theta-a}),
where the Laplace operator is substituted by its 3D radial
component. The boundary condition is
$\theta(r=R) =0$. The analysis
of the linear regime is similar to that for 1D and 2D cases and
gives for $t_D= R^2/D  \ll t \ll  \hbar/ \Delta $:
\begin{equation}
A= \pi^2 t/t_D, \qquad G(t) \sim \exp (- \pi^2 t/t_D)
\label{answer-3d-linear}
\end{equation}

The nonlinear in $\theta$ regime leads to the equation
\begin{equation}
\frac{d^2 \theta}{d z^2} + \frac{2}{z} \frac{d \theta}{d z}
+ i \omega t_D \sinh \theta =0, \quad z= \frac{r}{R}
\label{saddle-point-3d}
\end{equation}
The analysis of this equation shows that the permitted values
of $\omega$ are larger than a certain value $\omega_0 >0$,
and that the integral in
Eq.~(\ref{self-cons-theta}) remains finite even for the solutions of
Eq.~(\ref{saddle-point-3d}) with $ \theta(r=0) \to \infty$.
As a result, the self-consistency equation cannot be satisfied for
sufficiently long time $ t \geq \hbar/\Delta$.
Thus,  for $|\omega t_D| \ll 1$,  all non-trivial
solutions of Eq.~(\ref{saddle-point-3d})  satisfying the
condition $\theta (1)=0$  are singular at $ z \to 0$. Therefore,
the derivative $ d \theta / d r $ becomes comparable with
 the inverse mean-free path $1/l$ for a certain radius $r_*$. The
diffusion approximation inevitably breaks down for smaller
distances, where  non-local corrections become important.

We do not try to solve the kinetic problem now  but assume  that the
mentioned non-locality smoothes out the singularity at the origin.  We
also assume that, similarly to what has happened in the 1D and 2D cases, the
nonlinear term in Eq.~(\ref{saddle-point-3d}) can be neglected at $ r >
r_* $ and is important for $ r \sim r_*$. Thus
\begin{eqnarray*}
&& \theta(r) \sim C ( \frac{R}{r} -1 ), \quad r > r_* \\
&& 1 = l \frac{d\theta}{d r} \left.\right|_{r=r_*} =
\frac{C l R}{r_*^2},
\end{eqnarray*}
and $r_* = (C l R)^{1/2}$. Then $ \theta_* = \theta(r_*) =
(CR/l)^{1/2}$ and, finally,
$$
\omega t_D \exp (\theta_*) \sim \frac{\theta_* R^2}{r_*^2}
\sim \frac{1}{\theta_*} \left( \frac{R}{l} \right)^2,
$$
which gives
$$ \theta_* = \log \left[ \frac{1}{\omega t_D} \left(
\frac{R}{l} \right)^2 \right]
,\;
C = \frac{l}{R} \log^2 \left[ \frac{1}{\omega t_D} \left(
\frac{R}{l} \right)^2 \right].
$$

Using the self-consistency condition  we
express the frequency $\omega$ through the time $t$
and obtain a rough estimate for the action
\begin{equation}
A \sim \left( \frac{l}{\lambda} \right)^2
\log^\alpha \left( \frac{t}{\tau} \right), \;
G(t) \sim \exp \left[ - A(t) \right],
\label{answer-3d}
\end{equation}
where $\lambda = \hbar /p_F$ is the Fermi wavelength and the exponent
$\alpha \sim 1$ can only be determined from the solution
of the kinetic problem. Equation~(\ref{answer-3d})
looks similar to the AKL result if $\alpha=2$. The action~(\ref{answer-3d})
becomes comparable with that from Eq.~(\ref{answer-3d-linear}) at
\begin{equation}
t = t_*= t_D \left( \frac{l}{\lambda}\right)^2 = \frac{\hbar}{\Delta}
\frac{l}{R},
\quad t_D \ll t_* \ll \frac{\hbar}{\Delta}
\end{equation}

Therefore, we have two independent contributions to the conductance $G(t)$,
which is dominated by one of them ( Eq.~(\ref{answer-3d-linear})) at $ t < t_*$
and by the other (Eq.~(\ref{answer-3d})) at $ t > t_*$.

It is sensible now to analyze whether the diffusion treatment is valid
in the 1D and 2D cases. Using the solutions of Eqs.~(\ref{saddle-point-1d})
and (\ref{saddle-point-2d}) we find the value of $t_*$ such that
for $ t < t_*$ the derivative $ l d \theta/ d r $ is less than unity.
This gives:
\begin{equation}
t_* = \frac{h}{\Delta} \left\{ \begin{array}{c} e^{L/l}, \quad  d=1
                                \\ \left(\frac{R}{l} \right)^2, \quad d=2
                                  \end{array}
                               \right.
\end{equation}
Therefore, we can expect that the asymptotes (\ref{answer-1d}) and
(\ref{answer-2d}) are valid for $ t < t_*$. At longer times $ t > t_*$
for all dimensions $d=1,2,3$
the asymptote cannot be found within the diffusion approximation.
A detailed kinetic analysis of this problem will be done in a separate work.

We want now to speculate on the possible relation of this long time tail to
the AKL asymptote. AKL studied the coefficient growth rate in a power
expansion of $G(\omega)$ in $ \omega \tau$.  The rate they  found
means that $G(t)$ has a universal logarithmically-normal asymptote at
very long times. The presence of the mean free time $\tau$ in this
expansion makes it plausible that the AKL asymptote is related to a kind
of kinetic problem. This is why we hope that
at $ t \gg t_*$ the tail will match the results
of Ref.~\cite{Alt-Kra-Ler}.

In the case of a tunnel barrier at the sample-lead interface ,
the time dispersion of the conductance can be considered in the same
way with the usage of the generalized boundary condition
(see Ref.\cite{Kupriyanov}).


\begin{references}
\bibitem{conductance fluctuations} Our results represent the
time-dependent conductance $G(t)$ of a large set of
mesoscopic junctions with identical macroscopic parameters.
For a single junction the fluctuations of the conductance $G(t)$
are small compared to its averaged value, if $t \Delta \ll 1$, and
are not small in the opposite limit.
\bibitem{prefactor} The explicite form of prefactor $P\{Q\}$  is
irrelevant  within the exponential accuracy.
\bibitem{orthogonal} To be specific, we consider the orthogonal ensemble.
As it follows from the derivation below, to the exponential
accuracy,  the results do not depend on the ensemble.
\bibitem{solution-2d-small} The substitutions $ z = e^{-\eta} $ and
$\theta = \psi(\eta) + 2 \eta $ transform
Eq.~(\ref{saddle-point-2d}) into the form $  \psi'' + \gamma^2/4
e^\psi =0$   which has the first integral.
\end{references}
\end{document}